\documentclass{aastex63}
\usepackage{longtable}

\newcommand\teff{$\mathrm{T_{eff}}$}
\newcommand\menv{$\mathrm{M_{env}}$ }
\newcommand\menvtwo{$\mathrm{M_{env2}}$ }

\newcommand\menvtwons{$\mathrm{M_{env2}}$}
\newcommand\mhe{$\mathrm{M_{he}}$ }
\newcommand\mhens{$\mathrm{M_{he}}$}
\newcommand\mh{$\mathrm{M_H}$ }
\newcommand\mhns{$\mathrm{M_H}$}

\newcommand\kuv{KUV03442+0719 }
\newcommand\kuvns{KUV03442+0719}
\newcommand\bvf{Brunt-V\"ais\"aila }
\newcommand\msun{$\mathrm{M_\odot}$}
\newcommand\logg{log$g$ }
\newcommand\sigrms{$\sigma_{\rm RMS}$}

\shorttitle{Asteroseismic study of \kuv}
\shortauthors{Agnes Kim}

\graphicspath{{./}{figures/}}

\begin{document}

\title{Asteroseismic study of \kuv with parallax constraints}

\author{Agnes Kim}
\affiliation{Penn State Scranton \\
120 Ridgeview Drive \\
Dunmore, PA 18512, USA}

\begin{abstract}
Hydrogen atmosphere white dwarf KUV03442+0719 was first reported as a pulsator by Gianninas et al. in 2006. Follow up campaigns by Su et al. (2014) revealed more periods. Some spectroscopic results suggest that KUV03442+0719 has a slightly below average mass and an effective temperature of ~11000 K. But Gaia data (parallax and magnitude) suggest that it may be a low mass white dwarf. Such an object would have a helium core. We perform the asteroseismic fitting of KUV03442+0719, modeling it both as a carbon/oxygen normal mass white dwarf, and a helium core, low mass white dwarf. To perform the study, we perfom a grid search with WDEC models, refined by simplex minimization of the best fits. Both analyses result in best fit models that are comparable in terms of quality of fit. More pulsation data would be required to allow us to distinguish between the two scenarios. We present and constrast our results with expectations from stellar evolution. We also provide analytic formulae for a temperature dependent mass-radius relationship for helium core white dwarfs.
\end{abstract}

\section{Introduction} \label{sec:intro}
The focus of this paper is the asteroseismic fitting of pulsating white dwarf \kuvns, a hydrogen atmosphere pulsating white dwarf (ZZ Ceti star or DAV). With the recent addition of DAVs discovered with the TESS mission \citep{Romero22}, there are 500 objects in this class. White dwarfs are the end result of the evolution of lower mass stars ($\sim$ 98\% of all stars) and as such, hold in their interiors the fossil records of the evolution of their progenitor. White dwarf asteroseismology allows us to determine the interior structure of pulsating white dwarfs. The resulting chemical profiles help constrain physical processes such as nuclear fusion, core overshooting, massloss and diffusion.

Pulsations observed in white dwarfs are driven in the convection zone and are $g$-modes. Because of geometric cancellations, we do not expect to observe modes past $\ell$=2 \citep{Dziembowski77}. Modes are also described by their radial overtone. In this paper, we shall use $k$ to denote that number, even though it is often called $n$ in the literature. Most observed pulsation spectrum consist of fewer than a dozen modes. Sometimes, rotationally split triplets allow us to identify modes as most likely $\ell=1$, while quintuplets positively point to $\ell$=2 modes. While it is often not possible to do, positive identification of the modes before attempting the asteroseismic fitting is desirable, as best fit solutions can change according to which mode identification one adopts. This is why having constraints beyond the pulsation spectrum is key. The first pulsations in \kuv were reported in \citet{Gianninas06}. \citet{Su14a} performed a 3 year observing campaign that yielded more periods.

For \kuvns, there are a number of spectroscopic studies (section \ref{sec:gaia}). Two of these studies point to an average mass hydrogen atmosphere white dwarf. But the third, based on Gaia data, suggests that \kuv may be a low mass white dwarf \citep[0.29 \msun,][]{GF21}. The Gaia parallax allows the determination of a radius, if one assumes that the source is a single object. Based on that radius, \kuv is larger than the average white dwarf and therefore has a lower mass. White dwarfs are considered low mass (LMWDs) below 0.45 \msun. A subclass of LMWDs are the extremely low mass white dwarfs (ELMS), with masses between 0.18 and 0.20 \msun. \kuv does not have a known companion. Because the Main Sequence lifetime of the low mass progenitors of low mass white dwarfs is longer than the age of our galaxy,  it is commonly hypothesized that such stars are the result of binary system evolution, where the white dwarf progenitor loses mass to its companion during the red giant phase \citep[and references therein]{Althaus13}. Surveys have indeed found ELM's by looking at binary systems \citep{Brown22}. However, work both on the observational front \citep{Kilic07} and on the theoretical front \citep{Justham10} have suggested that a significant fraction of low mass white dwarfs could be found outside of binary systems. Formation mechanisms for such single objects include massive episodes of mass loss during the red giant phase due to high metalicities, or formation in wider binaries, where the white dwarf subsequently gets separated from its companion. Regardless of the formation mechanism, LMWDs are expected to have cores made up of helium instead of carbon and oxygen \citep{Althaus17}.

Another hypothesis is that \kuv is a member of an unresolved binary system (or more generally, that there is a line of sight object caught in the aperture). In that case, we may be looking at a normal mass, carbon and oxygen core white dwarf. While one can argue that this second scenario is more likely, the first scenario cannot be discarded based upon the available scientific evidence. Since it is not possible to determine which of these hypotheses is the right one, we proceed with studying \kuv under each hypothesis. One where we treat it as a helium core white dwarf and use constraints from \citet{GF21}, and one where we treat it as a carbon/oxygen core white dwarf and use constraints from \citet{Gianninas11} and \citet{Koester09}. In the latter case, we are unable to use distance constraints from Gaia, as the Gaia magnitude is that of the combined object. We begin this paper by revisiting the period analysis of \citep{Su14a} to produce a list of periods to use in the asteroseismic fitting. We then introduce and process the non-pulsational observational constraints we have for the object. In section \ref{sec:fitting} we introduce our fitting methods. We then proceed with the asteroseismic fitting of \kuv under the assumption of different core make-ups. We conclude in section \ref{sec:conclusions}.

\section{Observational constraints}
\label{sec:observations}

\subsection{Pulsation spectrum}
\label{sec:pulspectrum}

While \citet{Su14a} extract 31 independent modes in their work, based upon observations collected in 2010, 2011, and 2012, we do not think that the Fourier transforms shown in the paper support such an extensive list of periods. The signal to noise and spectral windows indicate that we can trust the two highest amplitude modes detected each year. We also check for beat frequencies and find none. The 6 frequencies resulting from this process are listed in table \ref{tab:periods1}. For ease of cross referencing with \citet{Su14a}, we use the same frequency labels. Frequencies with labels that start with "0" are present in the 2010 light curve, while the frequencies with labels that start with "1" and "2" are present in the 2011 and 2012 light curve respectively. The fact that the modes are sorted by frequencies is a coincidence. We also include the period found in the discovery paper \citep{Gianninas06}.

It is an established fact that the higher period modes in white dwarf interact with the base of the convection zone. This would be particularly true of \kuvns, as it is a red edge pulsator. \citet{Montgomery20} have quantified the effect on the frequencies. Higher $k$ modes can vary in frequency by as much as 5 $\mu$Hz. In table \ref{tab:periods1}, mode number 116 and 117 are less than 5 $\mu$Hz apart. The variations observed by \citet{Montgomery20} were measured over a time period of 75 days, while the 2011 observing run that gave rise to frequencies 116 and 117 was 7 nights long. However, given that the convective response timescales of white dwarfs are of the order of 100 seconds \citep{Provencal12} it is possible that 116 and 117 correspond to the same mode. Out of the pair, we include only mode 117 in our fitting, because of its much smaller error.

\begin{table}
  \begin{center}
  \caption{Period lists for \kuv. All are assumed to be $m=0$ modes. For the BIC parameters, $n_{obs} = 6$. For the C/O models, $n_{par} = 5$, while for the He model, $n_{par} = 6$ (based on the refined fitting, where \menv was fixed).
     \label{tab:periods1}
}
 {\scriptsize
  \begin{tabular}{|ccccc|cccccc|ccc|}
  \hline 
          &            &       &            &            & \multicolumn{6}{c|}{C/O model} & \multicolumn{3}{c|}{He model} \\
    Label & Frequency  & Error & Amplitude  & Observed   & \multicolumn{3}{c}{Global best fit} & \multicolumn{3}{c|}{Constrained $X_O$}  & & & \\
          & ($\mu Hz$) &       & ($mma$)    & periods (s) & Periods & $\ell$ & $k$ &  Periods & $\ell$ & $k$ & Periods & $\ell$ & $k$\\
    \hline
    006             & 741.17 & 0.20 & 8.81  & 1349.22   & 1348.368  & 1  & 23 & 1355.276 & 2  & 43 & 1349.926 & 1  & 15 \\
    007             & 755.67 & 0.20 & 9.80  & 1323.33   & 1323.217  & 2  & 40 & 1324.777 & 1  & 24 & 1327.360 & 2  & 27 \\
    116             & 777.00 & 0.28 & 5.53  & 1287.00   & na        & na & na & na       & na & na & na       & na & na \\
    117             & 780.12 & 0.05 & 6.72  & 1281.85   & 1281.779  & 1  & 22 & 1281.631 & 1  & 23 & 1281.807 & 1  & 14 \\
    205             & 834.21 & 0.13 & 2.73  & 1198.74   & 1198.216  & 2  & 36 & 1196.977 & 2  & 38 & 1199.120 & 2  & 24 \\
    206             & 843.44 & 0.19 & 2.60  & 1185.62   & 1185.025  & 1  & 20 & 1186.572 & 1  & 21 & 1187.457 & 1  & 13 \\
    Gianninas 2006  & 722.07 & na   & 7.6   & 1384.9    & 1385.015  & 2  & 42 & 1382.982 & 2  & 44 & 1380.884 & 2  & 28 \\
    \hline
                    &        &      &       & \sigrms       & 0.43~s &   &    & 1.31~s   &    &    & 1.11~s   &    &    \\
                    &        &      &       & $BIC_{K\& L}$ & -0.077 &   &    & 0.88     &    &    & 0.87     &    &    \\
                    &        &      &       & $BIC_{Liddle}$& -1.07  &   &    & 12.2     &    &    & 12.0     &    &    \\
    \hline
 \end{tabular}
 }
 \end{center}
\vspace{1mm}
\end{table}

\subsection{Spectroscopy, photometry and parallax}
\label{sec:gaia}

We list in table \ref{tab:properties} the effective temperature and surface gravity determinations that have been published for the star. Three results were obtained using solely spectroscopic modeling. They point to a white dwarf of average mass (\logg $\sim$ 8). The fourth surface gravity determination combines the spectroscopic measurement of the effective temperature, the Gaia (G) magnitude of the object and its parallax. It is brighter than an average mass white dwarf would be at that distance, indicating a larger than normal radius. This in turns leads to a lower surface gravity.

This points to two possible scenarios. In the first scenario, there is a line of sight companion to \kuv that augments its brightness. In performing the spectroscopic study of the object, we assume that the hydrogen lines used in the spectroscopic analysis belong to the pulsating white dwarf. Given the way the high surface gravities of white dwarfs shape their hydrogen lines, this does not seem completely unreasonable, but we have have to recognize that we are making an assumption here, when leveraging the \logg and effective temperature to constrain our asteroseismic fitting.

In the second (less likely) scenario, we are looking at a single object, and it must be a low mass, helium core white dwarf. If we assume that the G magnitude is entirely due to the white dwarf, then the photometry and the Gaia parallax allow us to place constraints on the mass and effective temperature of our best fit(s). This requires the use of a mass-radius relationship. We derive one from our own grid of helium core models (see section \ref{sec:fitting}), allowing us to obtain constraints that are self-consistent. In the method below, we take as input solely photometry and the parallax for \kuvns, with error bars. Those data are listed in the top three rows of table \ref{tab:properties}.

\begin{table}[!htbp]
\caption{Parameters for \kuv used in the mass determination from the Gaia parallax. Temperatures and \logg ~ from $^a$ \citet{GF21}, $^b$ \citet{Gianninas11}, $^c$ \citet{Gianninas11} with 3D corrections (Kepler, private communication), $^d$ \citet{Koester09}
  \label{tab:properties}
}
\centering
       \begin{tabular}{ll}
    \toprule
     Distance               &   $139.19 \pm 1$ pc       \\
     G magnitude            &   $16.6 \pm 0.003$        \\
     $E(B-V)^a$             &   $0.064 \pm 0.0348$      \\
     $\mathrm{T_{eff}^a}$   &   $10308.91 \pm 125.41$ K \\
     $\mathrm{T_{eff}^b}$   &   $11180 \pm 173$ K       \\
     $\mathrm{T_{eff}^c}$   &   $10870 \pm 173$ K       \\
     $\mathrm{T_{eff}^d}$   &   $10474 \pm 14$ K        \\
     \logg$^a$              &   $7.184425 \pm  0.033058$\\
     \logg$^b$              &   $7.92 \pm 0.06$         \\
     \logg$^c$              &   $7.78 \pm 0.06$         \\
     \logg$^d$              &   $7.75 \pm 0.01$         \\
     \hline
     \hline
  \end{tabular}
\end{table}

From Stefan-Boltzman's law, we have

\begin{equation}
\label{eqn:stephboltz}
    \frac{R_{WD}}{R_\odot}=\left(\frac{T_\odot}{T_{WD}}\right)^2\left(\frac{L_{WD}}{L_\odot}\right)^{1/2},
\end{equation}
where

\begin{equation}
\label{eqn:eq2}
    \frac{L_{WD}}{L_\odot}=\left(\frac{d_{WD}}{d_\odot}\right)^{2}\left(\frac{F_{WD}}{F_\odot}\right).
\end{equation}

In equation \ref{eqn:eq2}, the distance to the white dwarf, $d_{WD}$ can be obtained from Gaia data, and the ratio of the fluxes $\left(\frac{F_{WD}}{F_\odot}\right)$ is obtained using the Gaia absolute magnitude of the white dwarf $G$ and the absolute magnitude of the Sun ($M=4.83$):

\begin{equation}
\label{eqn:eq3}
    \left(\frac{F_{WD}}{F_\odot}\right)=10^{\frac{M-G_{\rm{corr}}}{2.5}},
\end{equation}

\noindent where $G_{\rm{corr}}$ is the Gaia magnitude corrected for reddening \citep{Casagrande18}.

\begin{equation}
\label{eqn:eq4}
    G_{\rm{corr}}=G-A_v=G-2.740E(B-V)
\end{equation}

A polynomial fit to WDEC models for helium core DA's yields the following mass-radius relationship:

\begin{equation}
\label{eq:massradiusrel}
    \frac{M_{WD}}{M_\odot}=a(T) (\log R)^3+b(T) (\log R)^2 + c(T) \log R + d(T) ,
\end{equation}

\noindent where the radius R is in centimeters. The parameters a, b, c, and d are themselves cubic functions of effective temperature. We discuss this mass-relationship further in the appendix, and supply values for the parameters in equation \ref{eq:massradiusrel}. In the appendix, we discuss the dependence of the mass-radius relation on the internal structure and effective temperature of the models. We find a strong dependence on the hydrogen envelope mass \menvtwons.

\begin{figure}[h!]
\epsscale{0.60}
\plotone{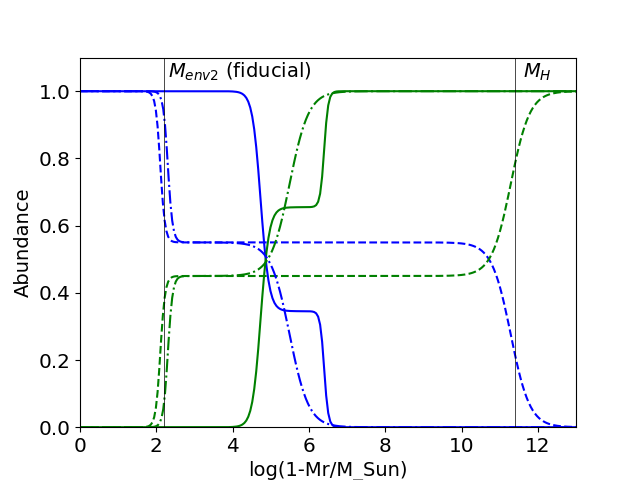}
\caption{Core profiles for the fiducial model that serves as a basis for the mass-radius relationship (dashed lines, section \ref{sec:observations}), best fit model (solid lines), and interior profile based on the 0.2724 \msun, 16481 K model of \citep{Calcaferro17} (dash-dotted line). The center is on the left. The core is composed of helium, while the envelope of hydrogen. The fiducial model, listed in table \ref{tab:hecoreparameters}, was chosen to have thick Helium/Hydrogen envelope, but a thin pure hydrogen layer. This allows maximum freedom in analyzing the dependence of the mass-radius relation on the different parameters. The best fit model is the result of the fitting procedure described in section \ref{sec:fitting}. \textbf{The labeled vertical lines illustrate how the values of the envelope parameters \menvtwo and \mh are defined. They are pictured for the fiducial model. }
\label{fig:hecomprofiles}}
\end{figure}

\begin{figure}[h!]
\epsscale{0.60}
\plotone{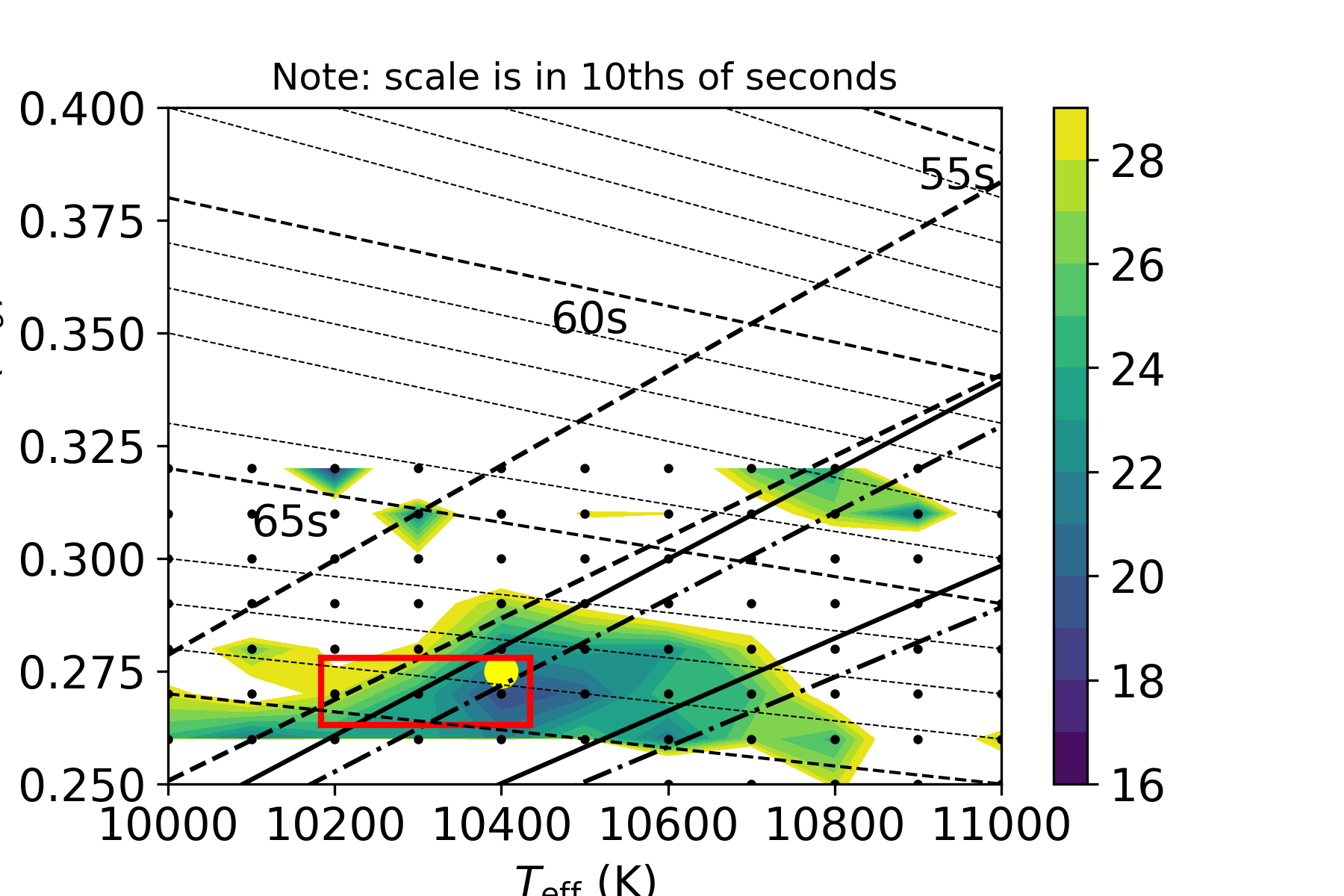}
\caption{Constraints for the asteroseismic fitting, along with a contour plot of the quality of fit of models comprising the master grid for helium core white dwarfs (see table \ref{tab:hecoreparameters}). The color scale is in tenths of seconds. The red box indicates the boundaries in effective temperature and mass based on the spectroscopy of \cite{GF21}. The diagonal lines with positive slope are constructed by combining the parallax data with the mass-radius relationship described in section \ref{sec:gaia}. The dashed lines correspond to a model with an envelope mass of \menvtwo = $10^{-2.2}$, the solid lines to \menvtwo = $10^{-5.0}$, and the dot-dashed lines to an envelope mass of $10^{-11.4}$. There is a pair of lines for each, a result of propagating the uncertainties in the distance and G magnitude. The closed circle marks the location of the best fit model (section \ref{sec:helium_core_fit}). We also plotted lines of asymptotic period spacings for l=1 modes calculated for our WDEC, He core white dwarf models (labeled every 5 seconds).
\label{fig:massteffconstraints}}
\end{figure}

Once we have the radius, we use the mass-radius relationship to find the corresponding mass. We thus obtain a relationship between the mass and the effective temperature. That relationship is plotted in Fig. \ref{fig:massteffconstraints} as the two bold solid diagonal lines.  We obtain two lines because we propagate the uncertainties on the distance and G magnitude. The dependence of the mass-radius relation on \menvtwo propagates to these lines, and we obtain different swaths for different values of \menvtwo (holding the other chemical structure parameters to that of the fiducial model).  

\subsection{Mode identification} \label{sec:modeid}

Since we do not have any clear triplets or quintuplets, we begin with the assumption that all observed modes are $m=0$ modes. All of the periods are greater than 1,100 seconds. For g-modes pulsations, this means that the period spectrum is close to the asymptotic limit and we expect little mode trapping (i.e. we expect the periods to be evenly spaced). This allows us to make use of the asymptotic period spacing to identify which modes are $\ell=1$ modes and which modes are $\ell=2$. 

In the theory of non-radial stellar oscillations, if we assume long periods (and so low frequencies), we find that the periods of the modes are given by \citep{Unno89}

\begin{equation}
\label{eqn:eq1}
    P_k=\frac{k\pi}{[ \ell(\ell+1)]^{1/2}}\left[ \int_{r_1}^{r^2}\frac{N}{r}dr\right]^{-1},
\end{equation}

\noindent where N is the \bvf frequency integrated between the turning points of the modes, and $k$ is the radial overtone of the mode. For a given star, the integral is a constant and we see that the periods are proportional to the factor $\frac{k}{\ell(\ell+1)^{1/2}}$. This means that there is a constant period spacing associated with each given $\ell$. We compute such average $\ell=1$ period spacings for a fiducial helium core model and show lines of constant period spacing in Fig. \ref{fig:massteffconstraints}. For carbon and oxygen core models of average mass ($\sim$ 0.60 \msun) and on the cool end of the ZZ Ceti instability strip, we find that the $\ell=1$ period spacing is $\sim$ 50~s, while for $\ell=2$ it is $\sim$ 30~s. With the period spacings as a guide, and given that we have only 6 periods to fit, it is not difficult to try every possible combination of $\ell=1$ and $\ell=2$ identifications and select the one that yields the best fits. We repeated the exercise for the helium core fitting. For the latter, Fig. \ref{fig:massteffconstraints} provides an idea of the expected $\ell=1$ period spacings. The $\ell=2$ spacing is a factor of $\sqrt{3}$ smaller.

\section{Fitting}
\label{sec:fitting}

Armed with the list of periods listed in table \ref{tab:periods1}, we proceed with the asteroseismic fitting. We calculated grids of models with the White Dwarf Evolution Code, WDEC. For details about the code see \citet{Bischoff-Kim18a}. The code is open source and may be obtained from GitHub (Codebase: https://github.com/kim554/wdec). A key feature of WDEC is the ability to vary the interior chemical profiles. Instead of using time dependent diffusion to calculate core chemical profiles based on some starting chemical composition, the code instead accepts the profiles as an input and then calculates a model that satisfies the equations of stellar structure and calculates the associated non-radial oscillation modes. The code allows to vary a maximum of 15 parameters: mass, effective temperature, convection, and 12 chemical profile parameters, described in \citet{Bischoff-Kim18a}. The details of the parameterization of the oxygen profile used in this work maybe be found in \citet{Bischoff-Kim18c}. 

For goodness of fit, we use the quantity
\begin{eqnarray}
\label{fiteq1}
\sigma_{\rm RMS} = \sqrt{\frac{1}{W} \sum_{1}^{n_{\rm obs}} {w_i(P^{\rm calc}_i-P^{\rm obs}_i)^2}}, \\
W=\frac{n_{\rm obs}-1}{n_{\rm obs}}\sum_{1}^{n_{\rm obs}}w_i
\end{eqnarray}

\noindent where $n_{\rm obs}$ is the number of periods present in the pulsation spectrum and the weights $w_i$ are the inverse square of the errors listed for each period in Table~\ref{tab:periods1}. For the period given in the discover paper \citep{Gianninas06}, we had to estimate a weight. We assigned the same weight as for modes 006 and 007, based on the similarities between the spectral windows and the fact that all are higher amplitude modes.

In order to help place the goodness of fits that we find for \kuv in context, it is useful to also provide a statistic called the "Bayes Information Criterion" (BIC). BIC's take into account the number of parameters versus the number of constraints (here periods) and give a measure of quality of fit that takes into consideration the fact that fewer periods fit with more parameters will lead to a smaller $\sigma_{RMS}$. Two BIC's have been used in the white dwarf asteroseismology literature, that of \citet{Koen00} and \citet{Liddle07}, equations \ref{eq:bickoen} and \ref{eq:bicliddle} respectively. The two quantities differ by a constant factor.

\begin{equation}
\label{eq:bickoen}
    BIC_{\rm K\&L}=\log(\sigma_{\rm RMS}^2)+ n_{\rm par} \left(\frac{\log n_{\rm obs}}{n_{\rm obs}}\right)
\end{equation}

\begin{equation}
\label{eq:bicliddle}
    BIC_{\rm Liddle}=n_{\rm obs}\ln(\sigma_{\rm RMS}^2)+ n_{\rm par} \ln n_{\rm obs}
\end{equation}

\subsection{Carbon oxygen core white dwarf fit} 
\label{sec:cocorefitting}

We begin with an assessment of the sensitivity of each available parameter to the period spectrum. This is a concern for \kuvns, as we only have 6 modes and because they all have higher periods, we expect a weak sensitivity to core structure. But it is worth checking, as it has been shown that there were exceptions to such rules \citep{Bischoff-Kim17a,Charpinet17}. 

\subsection{Parameter selection} 
\label{sec:parselection}

In order to quantify the influence each parameter has on the fits, we selected a fiducial model in an area of parameter space that is close to where we expect \kuv to land. For this guess, we used the spectroscopy of \citeauthor{Gianninas11} with 3D corrections (see table \ref{tab:properties}), and stellar evolution models by \citet{Althaus10}. Most notably, we adopted thick helium and hydrogen envelopes. At this stage, it is not crucial to have the best fit model for the star, but we do need to be in the right ballpark. The 15 parameters of the fiducial model are listed in table \ref{tab:cocoreparameters}. 

\begin{figure}[h!]
\plotone{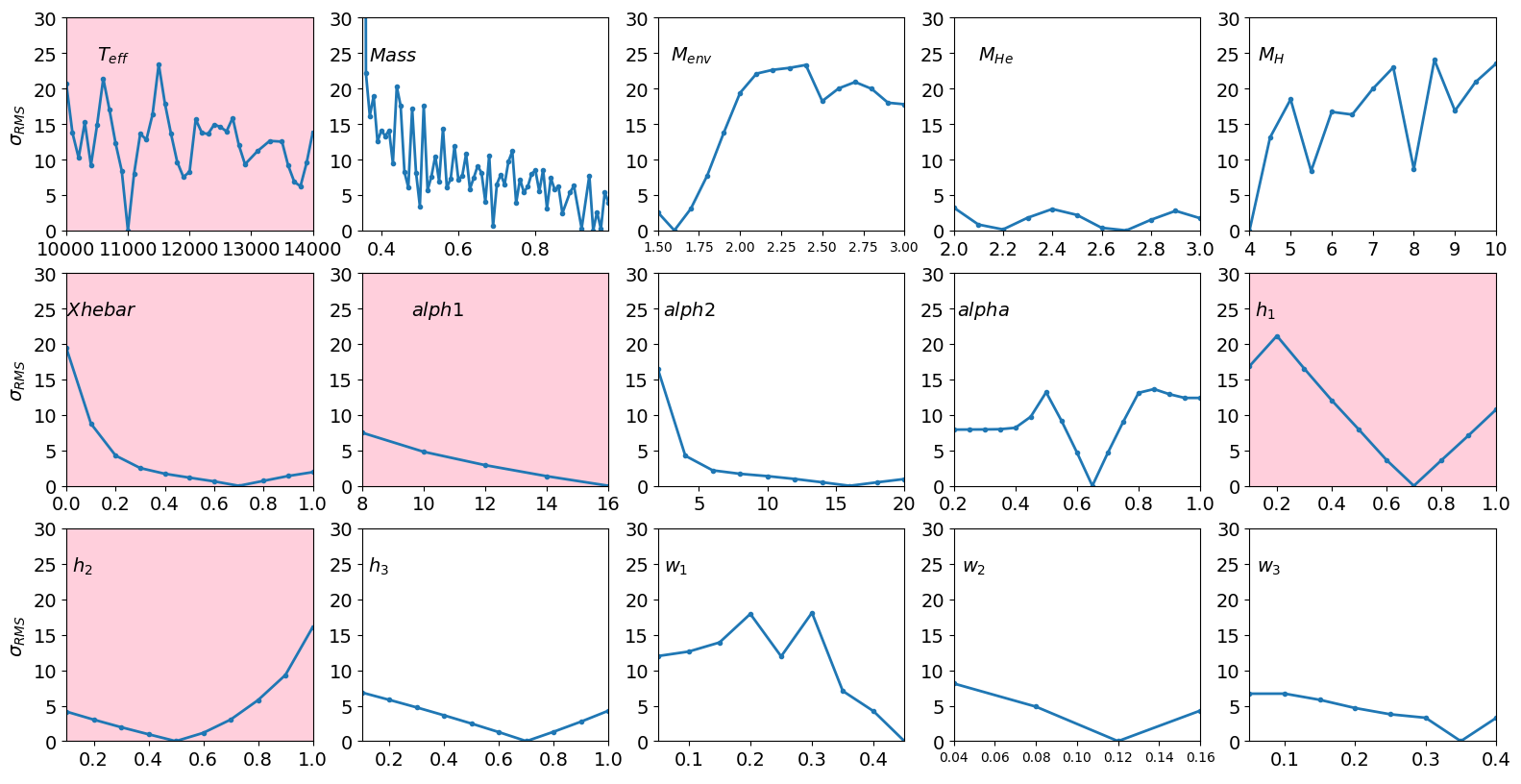}
\caption{Parameter sensitivity for \kuv's pulsation spectrum. For a description of the parameters, see text and \citet{Bischoff-Kim18c}. The lightly colored boxes highlight parameters that we varied in the grid search. \label{fig:pardep}}
\end{figure}

For that fiducial model, we varied one parameter at a time and observed the effect of varying that parameter on the quality of the fit. We tried his exercise for every possible $\ell$ identifications of the modes (this yielded 16 sets of periods to fit). The results for one representative set are shown in Fig. \ref{fig:pardep}. We find consistent results regardless of $\ell$ identifications. A perfect fit has a $\sigma_{\rm RMS}$ of zero seconds. Parameters \menv through alph2 dictate the shape of the helium and hydrogen composition profiles (they are envelope parameters). \menv is defined such that \menv$=10^{-2}$ separates the outer 0.01 (1\%) of the model where the helium and hydrogen reside from the inner 99\%, where only carbon and oxygen are present. \mhe and \mh are defined the same way, with the former marking the location of the base of the pure helium layer and the latter the location of the base of the hydrogen layer. alph1 and alph2 set how gradual the transitions are. The transition from pure helium to pure hydrogen is not parameterized, but instead calculated according to diffusive equilibrium. The parameter called "alpha" sets the strength of MLT convection, while the remaining parameters dictate the shape of the oxygen composition profile \citep{Bischoff-Kim18c}.

We can classify the behaviors in three categories: 1) parameters that strongly affect the quality of the fit and lead to well defined best fit values (for example the effective temperature and the envelope mass), 2) parameters that do not affect the quality of the fit at a significant level (for example \mhe and alph2), and 3) parameters that affect the fit in an unruly fashion, such as the stellar mass. As noted earlier, because of the long period modes present in \kuvns's spectrum, convection does affect the quality of the fits.

In figure \ref{fig:pardep}, we highlighted with color the parameters we ultimately decided to vary in the fits. While stellar mass has a significant effect on the quality of the fits, there is no monotonous descent to any given minimum and so we did not think we would learn much by varying that parameter. While \menv shows a clear minimum, its value is constrained by the thickness of the hydrogen layer. Since \mh falls in category 3 (unruly), we decided to fix that parameter to a canonical \mh = $10^{-4}$. \menv cannot be any smaller than $10^{-2}$ if \mh = $10^{-4}$. We fixed it to what stellar evolution calculations predict, around $10^{-1.6}$. The last envelope parameter, \mhens, does not affect the fits to a significant level. On the flip side, we find that even though we only have long period modes in the pulsation spectrum, they do affect parameters that set chemical profiles deeper in the interior (Xhebar, $h_1$ and $h_2$). WDEC can either treat convection as a free parameters, or use state of the art models to calculate it. We opted for the latter and did not treat that as a free parameter. We detail the parameters of the grid for CO core white dwarfs in table \ref{tab:cocoreparameters}.

\begin{table}
  \begin{center}
  \caption{Parameters used in the fits for the grid of CO core models. The values listed for \menv are $-\log({\rm M_{env}})$. Similarly for \mhe and \mhns.
     \label{tab:cocoreparameters}
     }
 {\scriptsize
  \begin{tabular}{lll}
  \hline 
	Oxygen Profile & Envelope Profiles & Other               \\
    \hline
	\multicolumn{3}{c}{Fiducial Model}                      \\
    $h_1=0.70$ & $M_{\rm env}=1.6$  & $T_{\rm eff}=11000$ K     \\
    $h_2=0.50$ & $M_{\rm He}=2.0$   & $M=0.465 M_\odot$            \\
    $h_3=0.70$ & xhe\_bar $=0.70$   & MLT $\alpha=0.65$          \\
    $w_1=0.45$ & ${\rm \alpha1}=16$ &                            \\
    $w_2=0.12$ & ${\rm \alpha2}=16$  &                            \\
    $w_3=0.35$ & $M_{\rm H}=4.0$    &                            \\
    \hline	               
	\multicolumn{3}{c}{Grid parameters - step sizes indicated after the semi-colon} \\
    $h_1=0.10,1.0$; 0.05  & \menv=1.6; fixed        & \teff=10100-11500;200 K         \\
    $h_2=0.10,1.0$; 0.10  & \mhe=2.0; fixed         & $M=0.465 M_\odot$; fixed          \\
    $h_3=0.70$; fixed  & xhe\_bar $=0.1,1.0$; 0.10  & MLT $\alpha$ calculated (see text)\\
    $w_1=0.45$; fixed  & ${\rm \alpha1}=4,20$; 2   &                                    \\
    $w_2=0.12$; fixed  & ${\rm \alpha2}=16$; fixed  &                                   \\
    $w_3=0.35$; fixed  & \mh=4.0; fixed             &                                   \\
    \hline
    \multicolumn{3}{c}{Parameters for best fit model 1}  \\
     $h_1=0.50$ & xhe\_bar = 0.10 & \teff = 10900 K\\
     $h_2=0.90$&  ${\rm \alpha1}=18$ & \\
     \hline
     \multicolumn{3}{c}{Parameters for best fit model 2}  \\
     $h_1=0.40$ &  xhe\_bar = 0.20 &\teff = 10900 K\\
     $h_2=0.90$ & ${\rm \alpha1}=20$& \\
     \hline
     \multicolumn{3}{c}{Parameters for best fit model 3}  \\
     $h_1=0.80$ &  xhe\_bar = 0.90 &\teff = 11100 K\\
     $h_2=1.00$ & ${\rm \alpha1}=4$& \\
     \hline
     \multicolumn{3}{c}{Global best fit}  \\
     $h_1=0.43$ &  xhe\_bar = 0.14 &\teff = 10905 K\\
     $h_2=0.92$ & ${\rm \alpha1}=20$& \\
     \hline
     \multicolumn{3}{c}{Constrained central oxygen abundance best fit}  \\
     $h_1=0.82$ &  xhe\_bar = 0.74 &\teff = 11133 K\\
     $h_2=0.98$ & ${\rm \alpha1}=4$& \\
    \hline
 \end{tabular}
  }
 \end{center}
\vspace{1mm}
\end{table}

\subsection{Helium core white dwarf fit} \label{sec:hefitting}

 A representative chemical composition profile for the helium core grid is shown in Fig. \ref{fig:hecomprofiles}. Five parameters describe the hydrogen profile. The location of the transition from pure helium to the He/H mix region is \menvtwons. The sharpness of that transition is described by the two diffusion parameters $\alpha_1$ and $\alpha_2$, with a higher value denoting a sharper transition. $\alpha_1$ cannot be much below 10, otherwise composition profiles show a discontinuity at the edge of the pure helium core for thicker hydrogen layers. The thickness of the pure hydrogen layer is given by the parameter \mhns. \textbf{In Fig. \ref{fig:hecomprofiles}, lines show the value of \menvtwo and \mh for the fiducial model}. \mh cannot be any smaller than $10^{-2.6}$, otherwise the transition zone is not smooth. There is also the constraint \mh $>$ \menvtwons. The hydrogen abundance in the mixed He/H region is denoted by $X_H$. To that we add mass, effective temperature and the efficiency of MLT convection \citep{Bohm71}. Because we do not have an oxygen profile for helium core white dwarfs, 6 parameters become irrelevant. We also no longer need the location of the base of the helium layer. This reduces the number of possible parameters to vary to 7, a computationally manageable number. Constraints in the mass-effective temperature plane from distances and magnitudes, described in section \ref{sec:gaia} help in the determination of a unique best fit. The parameters, the symbols used, and the range and step sizes used for the grids are listed in table \ref{tab:hecoreparameters}.

\begin{table}
  \begin{center}
  \caption{Parameters used in the fits for the grid of He core models. The values listed for \menvtwo are $-\log({\rm M_{env2}})$. Similarly for \mhns.
     \label{tab:hecoreparameters}
     }
 {\scriptsize
  \begin{tabular}{ll}
  \hline 
    Hydrogen profile                & Other                                                             \\
  \hline
    \multicolumn{2}{c}{Fiducial model}                                                                  \\
    \menvtwo=2.2                       & Varying effective temperature                                     \\
    \mh=11.4                         & Varying mass                                                      \\
    $X_H$=0.45                      & MLT $\alpha=0.65$                                                 \\
    ${\rm \alpha1}=16$              &                                                                   \\
    ${\rm \alpha2}=4$               &                                                                   \\
    \hline
    \multicolumn{2}{c}{Master grid parameters - step sizes indicated after the semi-colon}              \\
    \menvtwo=1.8,6.0; 0.8              & \teff=10000,11000; 100 K                                          \\
    \mh=max(\menvtwons,2.6),10.6; 0.8    & $M=0.26,0.32$; 0.01 $M_\odot$                                     \\
    $X_H$=0.,1.0; 0.2               & MLT $\alpha=0.6,0.8$; 0.2                                         \\
    ${\rm \alpha1}=10,16$; 2        &                                                                   \\
    ${\rm \alpha2}=4, 16$; 2        &                                                                   \\
    \hline
    \multicolumn{2}{c}{Refined grid parameters - step sizes indicated after the semi-colon}             \\
    \menvtwo=4.6,8.2; 0.4              & \teff=10300-10500; 50 K                                           \\
    \mh=5.0; fixed                  & $M=0.26,0.28; 0.005 M_\odot$                                       \\
    $X_H$=0.1,1.0; 0.05              & MLT $\alpha=0.8$; fixed                                           \\
    ${\rm \alpha1}=9,13$; 1        &                                                                   \\
    ${\rm \alpha2}=3,17$; 1        &                                                                   \\
    \hline
    \multicolumn{2}{c}{Best fit model}                                                \\
    \menvtwo=4.84                       & \teff=10438~K                                                     \\
    \mh=6.51                         & $M=0.270$ \msun                                                   \\
    $X_H$=0.65                       & MLT $\alpha=0.8$                                                  \\
    ${\rm \alpha1}=9.1$              &                                                                   \\
    ${\rm \alpha2}=18.7$             &                                                                   \\
   \hline
 \end{tabular}
  }
 \end{center}

\vspace{1mm}
\end{table}

\section{Results and discussion}
\label{sec:results}

\subsection{Carbon and Oxygen core}
\label{coresults}

\begin{figure}[h!]
\epsscale{1.15}
\plotone{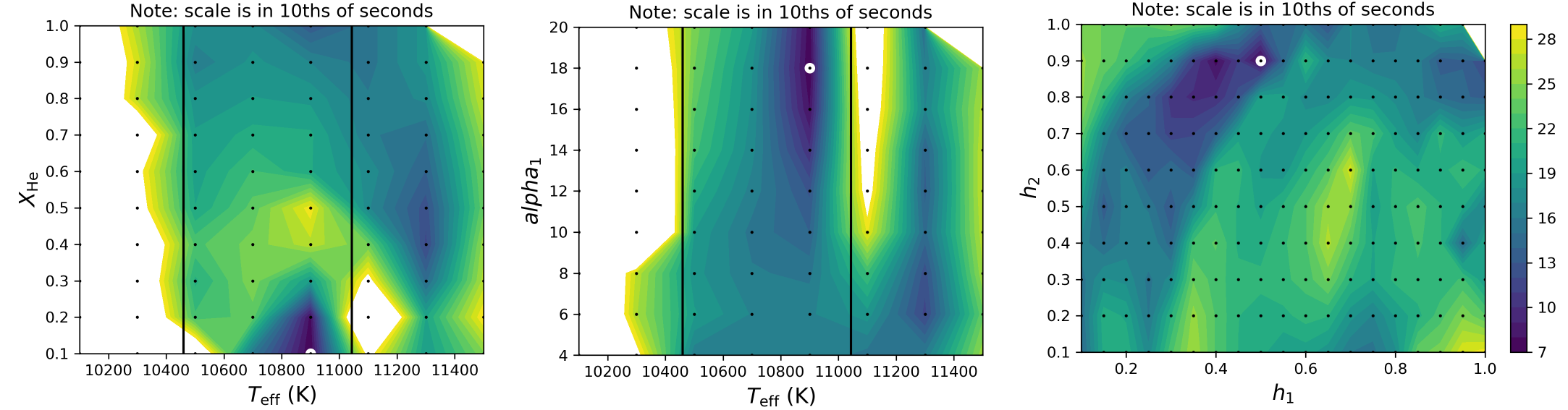}
\caption{Location of the best fits in three cuts in parameter space for C/O model 1. All five parameters varied in the C/O core fitting are featured. C/O model 2 present very similar contour plots. The vertical lines indicate the range of effective temperatures that correspond to the spectroscopy (\ref{tab:properties}). The third graph was produced using only the models that had effective temperatures in the spectroscopic range. 
\label{fig:co_contourplots}}
\end{figure}

\begin{figure}[h!]
\epsscale{0.6}
\plotone{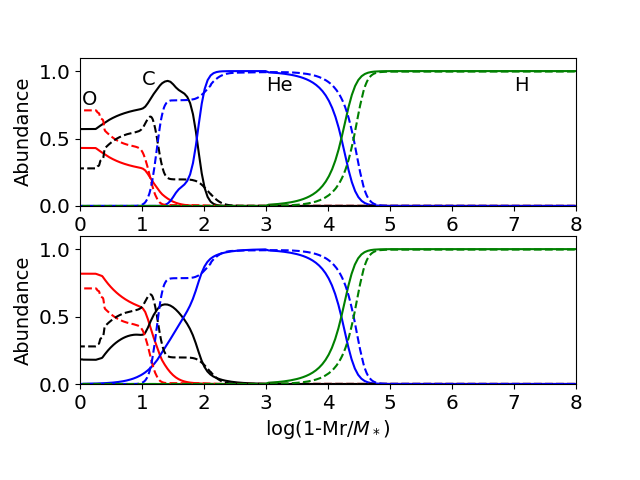}
\caption{Chemical composition profiles for the global best fit model of table \ref{tab:cocoreparameters} (solid lines, top panel) and for the best fit model constrained to have a central oxygen abundance greater than 0.50 (solid lines, bottom panel). The center of the model is on the left. For each, we also graph chemical profiles from \citet{Althaus10}. The models compare very closely in effective temperature, not in mass. The best fit models have a mass of 0.465 \msun, while the comparison models have a mass of 0.525 \msun. 
\label{fig:co_profiles}}
\end{figure}

Among all possible $\ell$ identification of modes, we find two clear best fit models. The parameters of these models (model 1 and model 2) are listed in table \ref{tab:cocoreparameters}. Model 1 is marginally better. In Fig. \ref{fig:co_contourplots} three contour maps that show the quality of the fits for that model. In choosing among all the possible $\ell$ identification we selected the combinations that yielded best fits that fell between 10,400~K and 11,100~K, to be consistent with the constraints from spectroscopy (table \ref{tab:properties}). The maps for C/O model 2 are very similar. The two best fit models differ by the $\ell$ identification of one of the higher error modes and land on identical or adjacent grid points for all parameter varied. We refined the best fit by performing a simplex search, using the $\ell$ identification of model 1. We graph the chemical profiles of C/O model 1 in the top panel Fig. \ref{fig:co_profiles}, along with chemical profiles for a 0.525 \msun, 10858~K model from \citet{Althaus10}, their lowest mass model. Model 1 has nearly the same effective temperature (10900~K), but a lower mass (0.465 \msun). 

Even if we cannot exactly compare the two models because of the mass discrepancy, a central oxygen abundance below 50\% is far below what is expected from stellar evolution calculations \citep{Corsico19}, especially for lower mass white dwarfs. That model also features a pure carbon layer between the C/O core and the He/H envelope, a feature that is very difficult to reproduce in stellar evolution calculations. If we constrain the central oxygen abundance to be greater than 50\%, we find a third good fit, labeled as "model 3" in table \ref{tab:cocoreparameters}. We also refined that best fit with a simplex search (Constrained central oxygen abundance best fit) in that same table. In table \ref{tab:periods1}, we list the periods of the two best fit models (the lower central oxygen abundance, global best fit, and the best fit constrained to a higher central oxygen abundance), with their $\ell$ and $k$ identifications. We also include BIC measures of quality of fit for each (eqns. \ref{eq:bickoen} and \ref{eq:bicliddle}). We graph the chemical profiles of the higher central oxygen abundance best fit model in Fig. \ref{fig:co_profiles} with a 0.525 \msun, 11359~K model from \citet{Althaus10}.

\subsection{Helium core}
\label{sec:helium_core_fit}
Because of the greater number of parameters, we started with a lower resolution grid, refined with a finer grid more narrowly focused on promising regions of parameter space. We started by fitting different period sets discussed in section \ref{sec:modeid} on the coarse grid. In order to choose the optimal period identification, we looked for best fit models that landed in or near the spectroscopic box shown in Fig. \ref{fig:massteffconstraints}. This gave us a constraint on the envelope mass (\menvtwo $\sim 10^{-5.0}$) helped us clearly identify the mode identification shown in table {\ref{tab:periods1}}. We refined the grid and finished the fitting with a simplex search to hone on the best fit parameters listed in table \ref{tab:hecoreparameters}. The corresponding chemical profiles are shown in Fig. \ref{fig:hecomprofiles}.

\section{Summary and conclusions}
\label{sec:conclusions}

We performed the asteroseismic fitting of \kuv, modeling it both as carbon-oxygen core, 0.465 \msun white dwarf, and as a low mass, helium core white dwarf. For the former, given Gaia parallax and magnitude data, we have to assume that there is a line of sight object that adds to the brightness of the white dwarf, while not leaving a signature in the spectrum. In the latter, the Gaia data is consistent with the lower mass (and therefore larger radius) of the white dwarf and we can assume that we are simply looking at the white dwarf. This allows us to use constraints from Gaia and from the spectroscopic temperature determination to limit our search in the mass-effective temperature plane. One useful product of this study is a mass-radius relationship for helium core white dwarfs, based on WDEC models. 

We contrast the interior chemical profiles of the best fit models we find for both the carbon-oxygen core white dwarfs and helium core white dwarfs with those that result from stellar evolution in Fig. \ref{fig:co_profiles} and \ref{fig:hecomprofiles} respectively. While we do not have entirely equivalent models, it is worth noting that for the helium core fit, we find a best fit model that has the expected hydrogen layer mass, but a thinner envelope mass. The carbon-oxygen model fits better than the helium core model, when we take into consideration the fact that we had more parameters varied in the latter, but only marginally so. To perform our study, we used periods published by \citet{Su14a}, but only selected the highest amplitude modes. Improved pulsation data might help better distinguish between the two scenarios in the future.

\section{Acknowledgments}
We thank Alejandro Corsico for an enlightening discussion of Bayes Information Criterions. S.O. Kepler and Keaton Bell offered independent opinions on the selection of the periods for the present study. This work was partly supported by NASA grant 80NSSC20K0455. It has made use of data from the European Space Agency (ESA) mission {\it Gaia} (\url{https://www.cosmos.esa.int/gaia}), processed by the {\it Gaia} Data Processing and Analysis Consortium (DPAC,
\url{https://www.cosmos.esa.int/web/gaia/dpac/consortium}). Funding for the DPAC has been provided by national institutions, in particular the institutions participating in the {\it Gaia} Multilateral Agreement. 

\software{WDEC \citep{Bischoff-Kim18b}}

\begin{appendix}

\begin{figure}[h!]
\epsscale{0.60}
\plotone{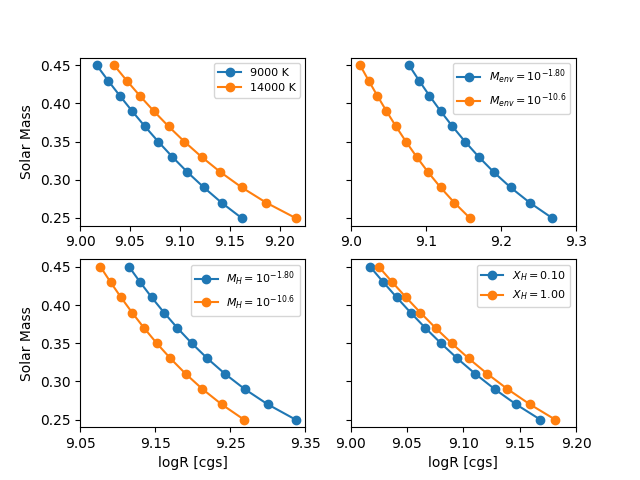}
\caption{Mass-radius relationships for different effective temperatures or structure parameters. The base model is the fiducial model listed in table \ref{tab:hecoreparameters}. The other parameters have a negligible effect on the mass-radius relation.
\label{fig:mass_radius_relations}}
\end{figure}

In this appendix, we expand on the analysis of the dependence of the mass-radius relation for helium core white dwarfs on the different parameters considered in this study and we provide coefficients for equation \ref{eq:massradiusrel} for models with different chemical structures. For this study, we use the fiducial model given in table \ref{tab:hecoreparameters} as a base. We vary each parameter one at a time and plot the mass-radius relation for extreme values of each. The results are shown in Fig. \ref{fig:mass_radius_relations}. We find that the envelope mass \menvtwo has the greatest effect on the mass-radius relation for helium core white dwarfs and so we take that effect into account in this work (in particular Fig. \ref{fig:massteffconstraints}. While we show the effect of the thickness of the pure hydrogen layer \mh for the full range of the parameter, it only differ from $10^{-11}$ and $10^{-5}$ between the fiducial and best fit model and so we expect that parameter to have a second order effect on our results.
 
\begin{table}
  \begin{center}
  \caption{Parameters for the mass-radius relation (Eqn. \ref{eq:massradiusrel} and subsequent text). The parameters are valid for $9,000~K<\rm{T_{eff}}<14,000~K$ and $0.25<\rm{M_*/M_\odot}<0.45$. They are given for different envelope masses. All other structure parameters are that of the fiducial model given in table \ref{tab:hecoreparameters}.
  \label{tab:massradiuspars_bestfit}
}
 {\scriptsize
  \begin{tabular}{ccccc}
    Coefficient & $T^3$ & $T^2$ & $T^1$ & $T^0$ \\
    \hline 
    \multicolumn{5}{c}{\menvtwo = $10^{-2.2}$} \\
    a &  -2.36592968e-11  &  8.76265483e-07 &  -1.11084137e-02 &  4.69379982e+01 \\
    b &   6.48133133e-10  & -2.39974997e-05 &   3.04118605e-01 & -1.28187851e+03 \\
    c &  -5.91780084e-09  &  2.19044467e-04 &  -2.77505404e+00 &  1.16670151e+04 \\
    d &   1.80093939e-08  & -6.66412494e-04 &   8.44006395e+00 & -3.53888806e+04 \\
    \hline 
    \multicolumn{5}{c}{\menvtwo = $10^{-3.4}$} \\
    a &  -2.88012742e-11  &  1.15535784e-06 &  -1.55858153e-02 &  6.94047379e+01 \\
    b &   7.86277483e-10  & -3.15406347e-05 &   4.25461171e-01 & -1.89151438e+03 \\
    c &  -7.15456332e-09  &  2.86992700e-04 &  -3.87113727e+00 &  1.71808595e+04 \\
    d &   2.16988624e-08  & -8.70404009e-04 &   1.17400108e+01 & -5.20109512e+04 \\
   \hline
   \multicolumn{5}{c}{\menvtwo = $10^{-5.0}$} \\
   a &  -6.58082570e-12  &  3.96808908e-07 &  -7.19574932e-03 &  4.01006970e+01 \\
   b &   1.77661482e-10  & -1.07570928e-05 &   1.95472628e-01 & -1.08752476e+03 \\
   c &  -1.59820514e-09  &  9.71865705e-05 &  -1.76978236e+00 &  9.82861519e+03 \\
   d &   4.79032697e-09  & -2.92616244e-04 &   5.34033256e+00 & -2.96005356e+04 \\
   \hline
   \multicolumn{5}{c}{\menvtwo = $10^{-6.6}$} \\
   a &   1.85090960e-11  & -4.97715229e-07 &   3.32336670e-03 & -4.91419588e-01 \\
   b &  -5.03499675e-10  &  1.35307230e-05 &  -9.01744776e-02 &  1.49528264e+01 \\
   c &   4.56596000e-09  & -1.22629213e-04 &   8.15778587e-01 & -1.52352609e+02 \\
   d &  -1.38035630e-08  &  3.70517614e-04 &  -2.46070061e+00 &  5.19009182e+02 \\
   \hline
   \multicolumn{5}{c}{\menvtwo = $10^{-11.4}$} \\
   a &  -2.34919591e-11  &  1.00499343e-06 &  -1.44771903e-02 &  6.94858832e+01 \\
   b &   6.40503430e-10  & -2.73981586e-05 &   3.94636848e-01 & -1.89080210e+03 \\
   c &  -5.82007802e-09  &  2.48940542e-04 &  -3.58537911e+00 &  1.71470677e+04 \\
   d &   1.76257118e-08  & -7.53859831e-04 &   1.08567730e+01 & -5.18237083e+04 \\
   \hline
  \end{tabular}
  }
 \end{center}
\vspace{1mm}
\end{table}

\end{appendix}

\bibliography{index}{}
\bibliographystyle{aasjournal}



\end{document}